\documentclass[12pt]{article}
\usepackage{amsmath,amssymb,amsthm,amsxtra,overpic,bbm,bm,epsfig,subfigure}
\usepackage{mathrsfs}
\usepackage{graphicx}
\usepackage{epstopdf}
\usepackage{color}
\usepackage{comment}
\usepackage{float}
\usepackage{cite}
\usepackage{slashed,stmaryrd}

\def\thefootnote{\fnsymbol{footnote}}

\begin{document}

\vspace{0.2cm}
\begin{center}
{\large\bf Naumov- and Toshev-like relations in the renormalization-group evolution of quarks and Dirac neutrinos}
\end{center}
\vspace{0.2cm}

\begin{center}
{\bf Zhi-zhong Xing~$^{a, b, c}$}~\footnote{E-mail: xingzz@ihep.ac.cn},
\quad
{\bf Shun Zhou~$^{a, b}$}~\footnote{E-mail: zhoush@ihep.ac.cn}
\\
{\small $^a$Institute of High Energy Physics, Chinese Academy of
Sciences, Beijing 100049, China \\
$^b$School of Physical Sciences, University of Chinese Academy of Sciences, Beijing 100049, China \\
$^c$Center for High Energy Physics, Peking University, Beijing 100871, China}
\end{center}

\vspace{1.5cm}

\begin{abstract}
In an analytical way of studying matter effects on neutrino oscillations, the Naumov and Toshev relations have been derived to respectively link the Jarlskog invariant of CP violation and the Dirac phase in the standard parametrization of the $3\times 3$ flavor mixing matrix to their matter-corrected counterparts. Here we show that there exist similar relations for Dirac neutrinos and charged leptons evolving with energy scales via the one-loop renormalization-group (RG) equations in the tau-dominance approximation, and for the running behaviors of up- and down-type quarks in the top-dominance approximation, provided a different parametrization is taken into account.
\end{abstract}

\newpage

\def\thefootnote{\arabic{footnote}}
\setcounter{footnote}{0}

\section{Introduction}

In particle physics the evolution of fermion masses and flavor mixing parameters from a superhigh energy scale to low scales (or vice versa) is described by the renormalization-group (RG) equations~\cite{Zhou}. Such a tool is extremely important because it helps bridge the gap between a predictive high-scale flavor model responsible for fermion mass generation and the relevant experimental data at low energies. Recently the authors of Refs.~\cite{Kuo} and \cite{XZZ} have borrowed the general idea of the RG language and applied it to the description of neutrino oscillations in matter. An interesting outcome of this successful application is the rediscovery of the Naumov relation~\cite{Naumov}
\begin{eqnarray}
\widetilde{\cal J}^{}_\ell \widetilde{\Delta}^{}_{21} \widetilde{\Delta}_{31} \widetilde{\Delta}^{}_{32} = {\cal J}^{}_\ell \Delta^{}_{21} \Delta^{}_{31} \Delta^{}_{32} \;
\end{eqnarray}
and the Toshev relation~\cite{Toshev}
\begin{eqnarray}
\sin \widetilde{\delta} \sin 2\widetilde{\theta}^{}_{23} =
\sin \delta \sin 2\theta^{}_{23} \; ,
\end{eqnarray}
which link the Jarlskog rephasing invariant ${\cal J}^{}_\ell$ of the $3\times 3$ Pontecorvo-Maki-Nakagawa-Sakata (PMNS) flavor mixing matrix $U$ and the Dirac CP-violating phase $\delta$ in the standard parametrization of $U$ \cite{PDG} to their matter-corrected counterparts $\widetilde{\cal J}^{}_\ell$ and $\widetilde{\delta}$, respectively. In the above equations $\Delta^{}_{ij} \equiv m^2_i - m^2_j$ (for $ij = 21, 31, 32$) denote the neutrino mass-squared differences in vacuum, $\theta^{}_{23}$ is one of the flavor mixing angles, and $\widetilde{\Delta}^{}_{ij}$ and
$\widetilde{\theta}^{}_{23}$ are the corresponding effective quantities
in matter. Note that the original derivations of Eq.~(1) \cite{Naumov} and Eq.~(2) \cite{Toshev} follow an {\it integral} approach, starting directly from the effective Hamiltonian responsible for neutrino oscillations in a medium
\footnote{In this spirit, the commutators or sum rules of charged-lepton and neutrino mass matrices can also be used to derive the Naumov relation \cite{Harrison,Xing2001,Xing2001x}.}.
In comparison, the new derivations done in Refs.~\cite{Kuo,XZZ} follow
the {\it differential} approach with the help of the RG-like equations.

The analogy between the RG evolution of neutrino masses and flavor mixing parameters with the energy scale and that with the matter parameter is so suggestive that we are wondering whether some interesting relations like Eqs. (1) and (2) in the latter case can also show up in the former case. Namely, we are eager to know whether there exist the Naumov- and Toshev-like relations for massive neutrino running from an arbitrary high energy scale $\mu$ down to the electroweak scale $\Lambda^{}_{\rm EW} \sim M^{}_Z = 91.2~{\rm GeV}$. At the one-loop level, we find that such relations {\it do} hold for Dirac neutrinos and charged leptons in the tau-dominance approximation, provided a different parametrization of the PMNS matrix $U$ advocated by Fritzsch and one of us~\cite{FX1,FX2} is taken into account
\footnote{Of course, only the Toshev-like relation is parametrization-dependent. A parametrization of the PMNS matrix $U$ which can make the tau-related elements much simpler will be favored in this regard. The matter effects on neutrino oscillations in an ordinary medium are only associated with the electron flavor, and hence the Toshev relation in Eq.~(2) can be derived when the standard parametrization of $U$, which makes the electron-related elements much simpler, is adopted.}.
Similar results are also obtainable for up- and down-type quarks in the top-dominance approximation. The main purpose of the present paper is just to report our findings and discuss their phenomenological implications.

\section{RG equations}

In the minimal extension of the standard model (SM) with three massive Dirac neutrinos, the RG evolution of the Yukawa coupling matrices for the up-type quarks $Y^{}_{\rm u}$, the down-type quarks $Y^{}_{\rm d}$, the Dirac neutrinos $Y^{}_\nu$ and the charged leptons $Y^{}_l$ at the one-loop level are governed by the following equations~\cite{XZ, Lindner:2005as}
\footnote{Here we do not consider the massive Majorana neutrinos, because their three CP-violating phases are entangled with one another during the RG evolution~\cite{XZ, Antusch:2003kp, Antusch:2005gp, Mei:2005qp}, making it impossible to obtain the concise Naumov- and Toshev-like relations which are only associated with the
Dirac CP-violating phase $\delta$.}
\begin{eqnarray}
16\pi^2 \frac{{\rm d} Y^{}_{\rm u}}{{\rm d}t} & = & \left[\alpha^{}_{\rm u} + \frac{3}{2} \left(Y^{}_{\rm u} Y^\dagger_{\rm u}\right) - \frac{3}{2} \left(Y^{}_{\rm d} Y^\dagger_{\rm d}\right)\right]Y^{}_{\rm u} \; ,
\nonumber \\
16\pi^2 \frac{{\rm d} Y^{}_{\rm d}}{{\rm d}t} & = & \left[\alpha^{}_{\rm d} - \frac{3}{2} \left(Y^{}_{\rm u} Y^\dagger_{\rm u}\right) + \frac{3}{2} \left(Y^{}_{\rm d} Y^\dagger_{\rm d}\right)\right]Y^{}_{\rm d} \; ,
\nonumber \\
16\pi^2 \frac{{\rm d} Y^{}_{\rm \nu}}{{\rm d}t} & = & \left[\alpha^{}_\nu
+ \frac{3}{2} \left(Y^{}_\nu Y^\dagger_\nu \right) - \frac{3}{2} \left(Y^{}_l Y^\dagger_l\right)\right]Y^{}_\nu \; ,
\nonumber \\
16\pi^2 \frac{{\rm d} Y^{}_l}{{\rm d}t} & = & \left[\alpha^{}_l
- \frac{3}{2} \left(Y^{}_\nu Y^\dagger_\nu\right) + \frac{3}{2} \left(Y^{}_l Y^\dagger_l\right)\right]Y^{}_l \; ,
\end{eqnarray}
where $t \equiv \ln(\mu/\Lambda^{}_{\rm EW})$ with $\mu$ being the renormalization energy scale and $\Lambda^{}_{\rm EW}$ being the electroweak energy scale, and
\begin{eqnarray}
\alpha^{}_{\rm u} &=& - \frac{17}{20} g^2_1 - \frac{9}{4} g^2_2 - 8 g^2_3 + \chi \; ,
\nonumber \\
\alpha^{}_{\rm d} &=& - \frac{1}{4} g^2_1 - \frac{9}{4} g^2_2 - 8 g^2_3 + \chi \; ,
\nonumber \\
\alpha^{}_{\nu} &=& - \frac{9}{20} g^2_1 - \frac{9}{4} g^2_2 + \chi \; ,
\nonumber \\
\alpha^{}_{l} &=& - \frac{9}{4} g^2_1 - \frac{9}{4} g^2_2 + \chi \; ,
\end{eqnarray}
with $\chi \equiv {\rm Tr}\left[3\left(Y^{}_{\rm u} Y^\dagger_{\rm u}\right) + 3 \left(Y^{}_{\rm d} Y^\dagger_{\rm d}\right) + \left(Y^{}_\nu Y^\dagger_\nu\right) + \left(Y^{}_l Y^\dagger_l\right)\right]$, and the ${\rm SU}(3)^{}_{\rm C} \times {\rm SU}(2)^{}_{\rm L} \times {\rm U}(1)^{}_{\rm Y}$ gauge couplings $g^{}_3$, $g^{}_2$ and $g^{}_1$ evolving via the RG
equations $16\pi^2 \left({\rm d}g^{}_i/{\rm d}t\right) = b^{}_i g^3_i$ with the coefficients $\left\{b^{}_3, b^{}_2, b^{}_1\right\} = \left\{-7, -19/6, 41/10\right\}$.

Without loss of generality, we shall work in the flavor basis where $Y^{}_{\rm u} = {\rm diag}\left\{y^{}_u, y^{}_c, y^{}_t\right\} \equiv D^{}_{\rm u}$ and $Y^{}_l = {\rm diag}\left\{y^{}_e, y^{}_\mu, y^{}_\tau\right\} \equiv D^{}_l$ are chosen. Given the fact that the top-quark mass is far above the bottom-quark mass and the other quark
masses, the $Y^{}_{\rm d} Y^\dagger_{\rm d}$ terms on the right-hand side of Eq.~(3) are safely negligible. It is certainly much safer to neglect the $Y^{}_\nu Y^\dagger_\nu$ terms, because the neutrino masses are far below the charged lepton masses. In these good approximations Eq.~(3) can be simplified to
\begin{eqnarray}
16\pi^2 \frac{{\rm d} D^{}_{\rm u}}{{\rm d}t} & = & \left[\alpha^{}_{\rm u} + \frac{3}{2} D^2_{\rm u}\right]D^{}_{\rm u} \; ,
\nonumber \\
16\pi^2 \frac{{\rm d} Y^{}_{\rm d}}{{\rm d}t} & = & \left[\alpha^{}_{\rm d} - \frac{3}{2} D^2_{\rm u} \right]Y^{}_{\rm d} \; ,
\end{eqnarray}
for quarks; and
\begin{eqnarray}
16\pi^2 \frac{{\rm d} Y^{}_{\rm \nu}}{{\rm d}t} & = & \left[\alpha^{}_\nu - \frac{3}{2} D^2_l \right]Y^{}_\nu \; ,
\nonumber \\
16\pi^2 \frac{{\rm d} D^{}_l}{{\rm d}t} & = & \left[\alpha^{}_l + \frac{3}{2} D^2_l\right]D^{}_l \; ,
\end{eqnarray}
for leptons, together with $\chi \simeq 3 y^2_t$. Then it is straightforward to formally integrate Eqs.~(5) and (6) from
$t = t^{}_0 \equiv \ln \left(\Lambda/\Lambda^{}_{\rm EW}\right)$
(e.g., a superhigh energy scale $\Lambda$ where there may exist a kind of
underlying flavor symmetry~\cite{Zhao}
\footnote{If massive neutrinos are Majorana particles, the superhigh energy scale $\Lambda$ can just be where the seesaw mechanism~\cite{Minkowski,Yanagida,Glashow,Gell-Mann,Mohapatra} works.})
down to $t = 0$ (i.e., the electroweak scale $\Lambda^{}_{\rm EW}$). In the chosen basis the Yukawa coupling matrices of up-type quarks ($D^{}_{\rm u}$) and charged leptons ($D^{}_l$) keep diagonal during the RG running.
To be more explicit, let us introduce the functions
\begin{eqnarray}
I^{}_{\rm f} = \exp\left[-\frac{1}{16\pi^2} \int^{t^{}_0}_0
\alpha^{}_{\rm f}(t) {\rm d}t\right] \; ,
\end{eqnarray}
for ${\rm f} = {\rm u}, {\rm d}, \nu, l$;
\begin{eqnarray}
\xi^{}_q = \exp\left[+\frac{3}{32\pi^2} \int^{t^{}_0}_0 y^2_q(t) {\rm d}t\right] \; ,
\end{eqnarray}
for $q = u, c, t$; and
\begin{eqnarray}
\zeta^{}_\alpha = \exp\left[+\frac{3}{32\pi^2} \int^{t^{}_0}_0 y^2_\alpha(t) {\rm d}t\right] \; ,
\end{eqnarray}
for $\alpha = e, \mu, \tau$. As a consequence, the fermion Yukawa coupling matrices at $\Lambda^{}_{\rm EW}$, which are denoted respectively as $D^\prime_{\rm u}$, $Y^\prime_{\rm d}$, $Y^\prime_\nu$ and $D^\prime_l$, can be expressed in an integrated way as
\begin{eqnarray}\label{eq:inteq}
D^\prime_{\rm u} = I^{}_{\rm u} T^{}_{\rm u} D^{}_{\rm u}\; , \quad
Y^\prime_{\rm d} = I^{}_{\rm d} T^{}_{\rm d} Y^{}_{\rm d} \; ,
\end{eqnarray}
and
\begin{eqnarray}\label{eq:intel}
D^\prime_l = I^{}_l T^{}_l D^{}_l \; , \quad
Y^\prime_\nu = I^{}_\nu T^{}_\nu Y^{}_\nu \; ,
\end{eqnarray}
where $T^{}_{\rm d} = T^{-1}_{\rm u} \equiv {\rm diag}\{ \xi^{}_u, \xi^{}_c, \xi^{}_t \}$ and $T^{}_{\nu} = T^{-1}_l \equiv {\rm diag}\{ \zeta^{}_e, \zeta^{}_\mu, \zeta^{}_\tau \}$. Note that the relationship between $Y^\prime_\nu$ and $Y^{}_\nu$ has previously been obtained in the minimal supersymmetric standard model (MSSM) and the two-Higgs-doublet model (2HDM) \cite{Xing:2017mkx}. Here we deal with all the four fermion sectors together in the SM framework.

In fact, $\xi^{}_u \simeq \xi^{}_c \simeq \zeta^{}_e \simeq \zeta^{}_\mu \simeq 1$ is an excellent approximation due to the smallness of $m^{}_u$, $m^{}_c$, $m^{}_e$ and $m^{}_\mu$, as compared with the vacuum expectation value of the Higgs field $v \simeq 174$ GeV. It is also reasonable to take $\zeta^{}_\tau \simeq 1$ in the SM, but we do not do so in the present paper. Fig. 1 illustrates the magnitudes of $I^{}_{\rm u}$, $I^{}_{\rm d}$, $I^{}_l$, $I^{}_\nu$, $\xi^{}_t$ and $\zeta^{}_\tau$
for different values of $\Lambda$, where the running quark and charged-lepton masses and other SM parameters renormalized to the energy scale $\mu = M^{}_Z$~\cite{Xing:2011aa} have been input. It is clear that $\xi^{}_t$ increases by $10\%$ when changing from $\Lambda^{}_{\rm EW} \sim M^{}_Z$ to $\Lambda = 10^{12}~{\rm GeV}$, whereas $\zeta^{}_\tau$ is almost unchanged. On the other hand, the values of $I^{}_{\rm u}$ and $I^{}_{\rm d}$ at the high-energy scale $\Lambda$ can be twice larger, while those of $I^{}_\nu$ and $I^{}_l$ can change by no more than $10\%$. Although the variation of $\zeta^{}_\tau$ is negligible in the SM, we should keep in mind that the Yukawa couplings of charged leptons in the MSSM are given by $y^2_\alpha = \left(1 + \tan^2\beta\right)m^2_\alpha/v^2$, which can be significantly enhanced for a relatively large value of $\tan\beta$. For this reason, we retain both $\xi^{}_t$ and $\zeta^{}_\tau$ in the following discussions, but take the approximation $\xi^{}_u \simeq \xi^{}_c \simeq \zeta^{}_e \simeq \zeta^{}_\mu \simeq 1$ whenever it is necessary.
\begin{figure}[t!]
\hspace{-1.0cm}
\subfigure{%
\includegraphics[width=0.6\textwidth]{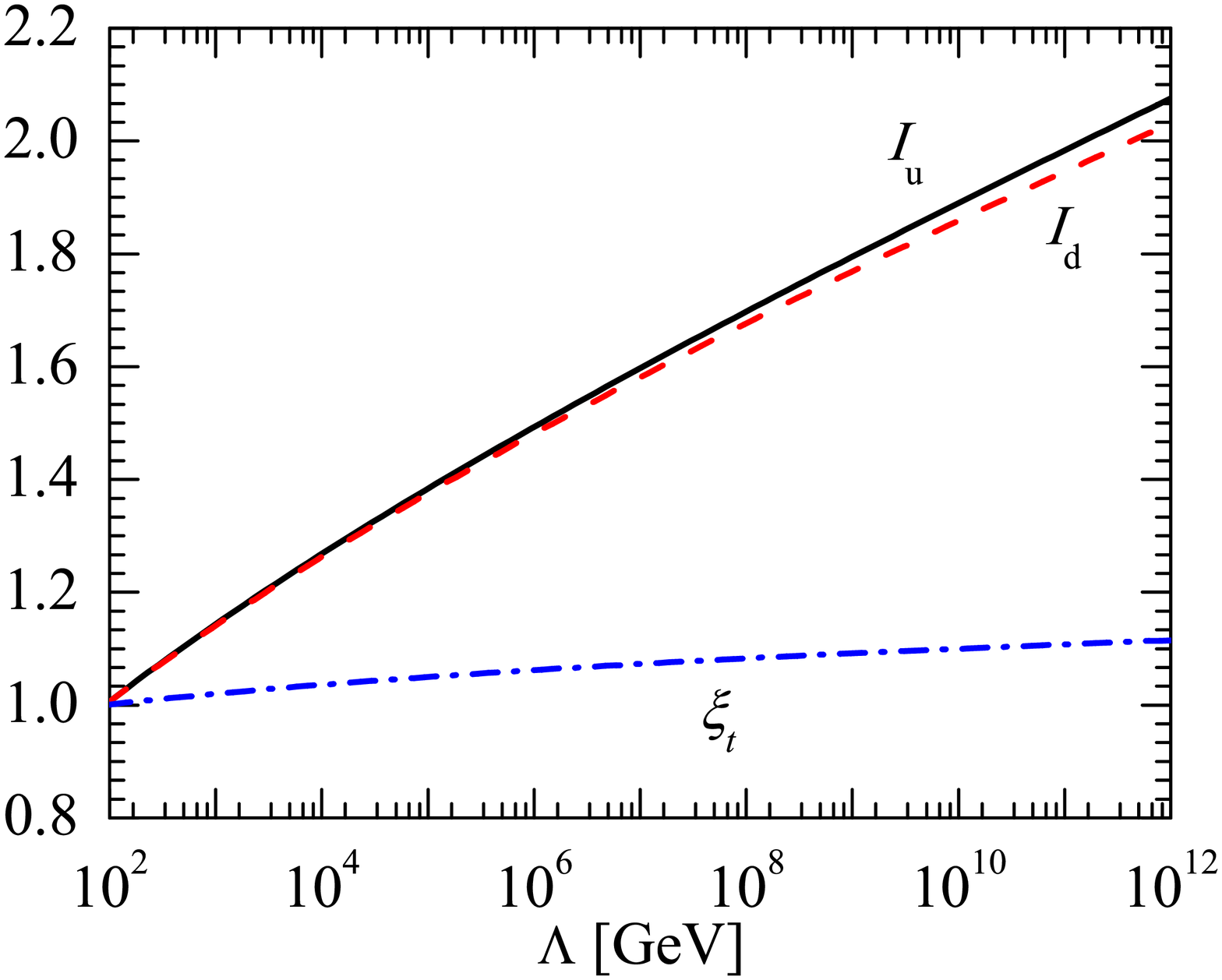}
 }%
\subfigure{%
\hspace{-1.7cm}
\includegraphics[width=0.6\textwidth]{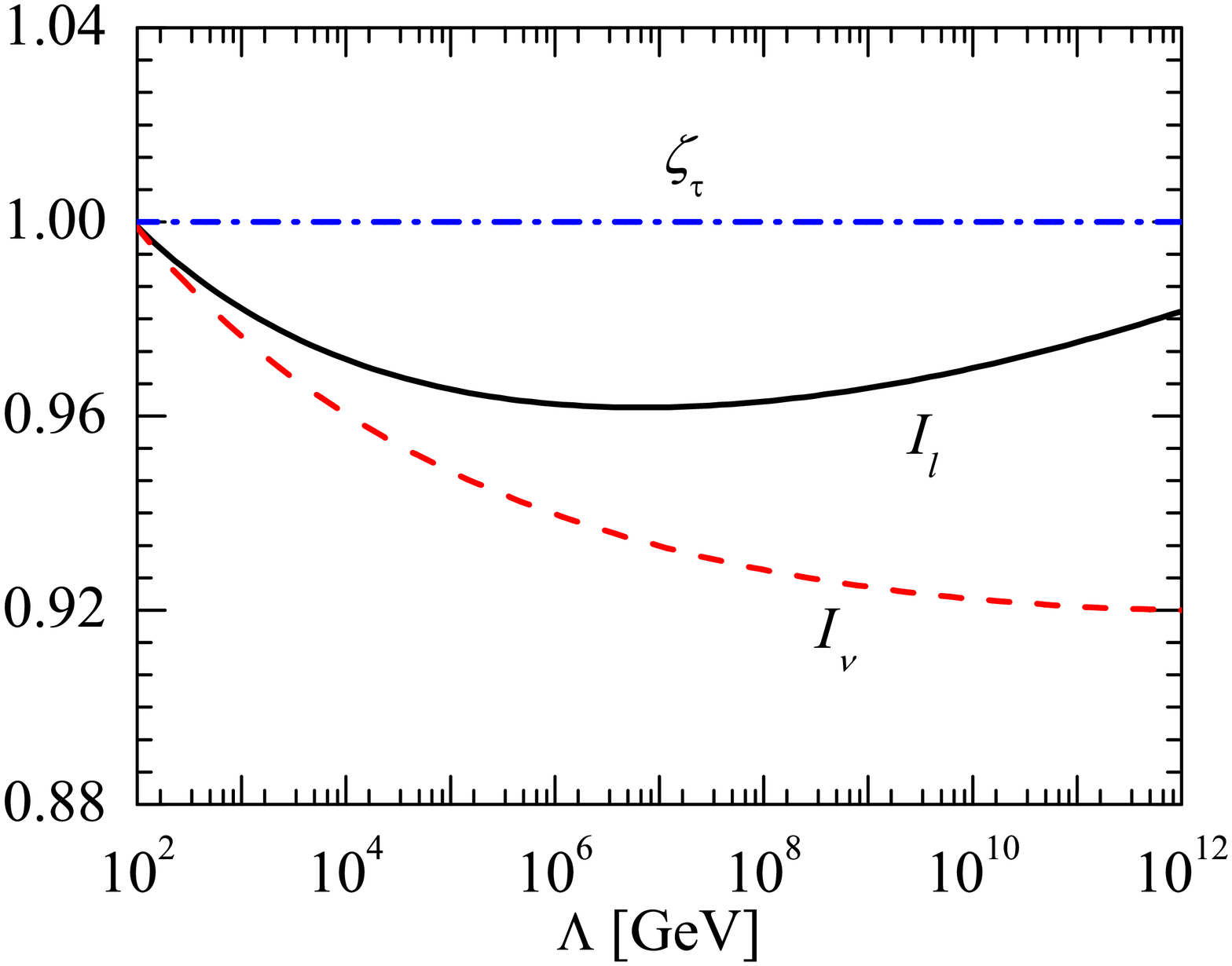}        }
\caption{Illustration for the evolution functions $I^{}_{\rm u}$, $I^{}_{\rm d}$ and $\xi^{}_t$ in the quark sector (left panel) and $I^{}_l$, $I^{}_\nu$ and $\zeta^{}_\tau$ in the lepton sector (right panel) for different values of $\Lambda$, where the running quark and charged-lepton masses and other SM parameters renormalized to the energy scale $\mu = M^{}_Z$ \cite{Xing:2011aa} have been input.}
\label{fig:fig1}
\end{figure}

\section{Naumov- and Toshev-like relations}

Given the relations between the fermion Yukawa coupling matrices at $\Lambda^{}_{\rm EW}$ and those at $\Lambda$ as established in Eqs.~(10) and (11), we now investigate their phenomenological implications and derive the Naumov-like relations for the RG evolution of quarks and Dirac neutrinos. In the chosen basis the Cabibbo-Kobayashi-Maskawa (CKM) quark flavor mixing matrix $V^\prime$ at $\mu = \Lambda^{}_{\rm EW}$ can be determined from the diagonalization of
\begin{eqnarray}
H^\prime_{\rm d} \equiv Y^\prime_{\rm d} Y^{\prime \dagger}_{\rm d} = V^\prime D^{\prime 2}_{\rm d} V^{\prime \dagger} \; ;
\end{eqnarray}
while the PMNS matrix $U^\prime$ at $\Lambda^{}_{\rm EW}$ is from diagonalizing
\begin{eqnarray}
H^\prime_\nu \equiv Y^\prime_\nu Y^{\prime \dagger}_\nu = U^\prime D^{\prime 2}_\nu U^{\prime \dagger}\; ,
\end{eqnarray}
in which $D^{\prime}_{\rm d} \equiv {\rm diag}\{y^{\prime}_d, y^{\prime}_s, y^{\prime}_b\} = {\rm diag}\{m^{\prime}_d, m^{\prime}_s, m^{\prime}_b\}/v$ with $m^{\prime}_q$ being the down-type quark mass (for $q = d, s, b$), and $D^{\prime}_\nu \equiv {\rm diag}\{y^{\prime}_1, y^{\prime}_2, y^{\prime}_3\} = {\rm diag}\{m^{\prime}_1, m^{\prime}_2, m^{\prime}_3\}/v$ with $m^{\prime}_i$ being the neutrino mass (for $i = 1, 2, 3$). At a superhigh-energy scale $\mu = \Lambda$ the CKM matrix $V$ and the PMNS matrix $U$ can be obtained in the same way as in Eqs.~(12) and (13).
The point is that both $D^{}_{\rm u}$ and $D^{}_l$ keep diagonal during the RG evolution, as
we have remarked before. So the eigenvalues of $H^{}_{\rm d} \equiv Y^{}_{\rm d} Y^\dagger_{\rm d}$ and $H^{}_\nu \equiv Y^{}_\nu Y^\dagger_\nu$ are given by $V^\dagger H^{}_{\rm d} V = D^2_{\rm d} \equiv {\rm diag}\{y^2_d, y^2_s, y^2_b\}$ and $U^\dagger H^{}_\nu U = D^2_\nu \equiv {\rm diag}\{y^2_1, y^2_2, y^2_3\}$, respectively.

Let us first focus on the commutators of up-type and down-type quark mass matrices at both low- and high-energy scales. More explicitly, we have
\begin{eqnarray}
\left[D^{\prime 2}_{\rm u}, H^\prime_{\rm d}\right] \equiv {\rm i} X^\prime_{\rm q} \; , \quad  \left[D^2_{\rm u}, H^{}_{\rm d}\right] \equiv {\rm i} X^{}_{\rm q}\; ,
\end{eqnarray}
where $X^\prime_{\rm q}$ and $X^{}_{\rm q}$ are obviously two Hermitian matrices and their expressions can be figured out by directly calculating the commutators, i.e.,
\begin{eqnarray}
X^\prime_{\rm q} & = & {\rm i} \left(\begin{matrix} 0 & \Delta^\prime_{cu} Z^\prime_{uc} & \Delta^\prime_{tu} Z^\prime_{ut} \cr \Delta^\prime_{uc} Z^\prime_{cu} & 0 & \Delta^\prime_{tc} Z^\prime_{ct} \cr \Delta^\prime_{ut} Z^\prime_{tu} & \Delta^\prime_{ct} Z^\prime_{tc} & 0\end{matrix}\right) \; ,
\nonumber \\
\quad X^{}_{\rm q} & = & {\rm i} \left(\begin{matrix} 0 & \Delta^{}_{cu} Z^{}_{uc} & \Delta^{}_{tu} Z^{}_{ut} \cr \Delta^{}_{uc} Z^{}_{cu} & 0 & \Delta^{}_{tc} Z^{}_{ct} \cr \Delta^{}_{ut} Z^{}_{tu} & \Delta^{}_{ct} Z^{}_{tc} & 0 \end{matrix}\right) \; ,
\end{eqnarray}
with the notations $\Delta^\prime_{pq} \equiv y^{\prime 2}_p - y^{\prime 2}_{q}$, $\Delta^{}_{pq} \equiv y^{2}_p - y^{2}_{q}$,  $Z^\prime_{p q} \equiv y^{\prime 2}_d V^{\prime}_{p d} V^{\prime *}_{q d} + y^{\prime 2}_s V^{\prime}_{p s} V^{\prime *}_{q s} + y^{\prime 2}_b V^{\prime}_{p b} V^{\prime *}_{q b}$ and $Z^{}_{p q} \equiv y^{2}_d V^{}_{p d} V^*_{q d} + y^{2}_s V^{}_{p s} V^*_{q s} + y^{2}_b V^{}_{p b} V^*_{q b}$ (for the subscripts $p$ and $q$ running over the flavors $u$, $c$ and $t$). It is straightforward to verify that the determinant of the above commutator leads us to a rephasing-invariant measure of CP violation~\cite{Jarlskog:1985ht}:
\begin{eqnarray}
{\rm det}\left(\left[D^{\prime 2}_{\rm u}, H^\prime_{\rm d}\right]\right) = 2{\rm i} \Delta^\prime_{cu} \Delta^\prime_{tc} \Delta^\prime_{tu} {\rm Im} \left[Z^\prime_{cu} Z^\prime_{tc} Z^\prime_{ut}\right] = 2{\rm i} \Delta^\prime_{cu} \Delta^\prime_{tc} \Delta^\prime_{tu} \Delta^\prime_{sd} \Delta^\prime_{bd} \Delta^\prime_{bs} {\cal J}^\prime_{\rm q} \; ,
\end{eqnarray}
in which ${\cal J}^{\prime}_{\rm q}$ is the Jarlskog invariant of the CKM matrix at $\Lambda^{}_{\rm EW}$. On the other hand, the identities in Eq.~(10) lead us to $H^\prime_{\rm d} = I^2_{\rm d}T^{}_{\rm d} H^{}_{\rm d} T^{}_{\rm d}$ and thus $[D^{\prime 2}_{\rm u}, H^\prime_{\rm d}] = I^2_{\rm d} T^{}_{\rm d} [D^{\prime 2}_{\rm u}, H^{}_{\rm d}] T^{}_{\rm d}$, where $\left[D^{\prime 2}_{\rm u}, T^{}_{\rm d}\right] = 0$ should be noticed. Calculating the determinant of the commutator along this line, we immediately arrive at
\begin{eqnarray}
{\rm det}\left(\left[D^{\prime 2}_{\rm u}, H^\prime_{\rm d}\right]\right) = {\rm det} \left(I^2_{\rm d} T^{}_{\rm d} \left[D^{\prime 2}_{\rm u}, H^{}_{\rm d}\right] T^{}_{\rm d} \right) = 2{\rm i} I^6_{\rm d} \xi^2_u \xi^2_c \xi^2_t \Delta^\prime_{cu} \Delta^\prime_{tc} \Delta^\prime_{tu} \Delta^{}_{sd} \Delta^{}_{bd} \Delta^{}_{bs} {\cal J}^{}_{\rm q} \; .
\end{eqnarray}
A combination of Eqs.~(16) and (17) can therefore give rise to the Naumov-like relation for the RG evolution of quarks:
\begin{eqnarray}
{\cal J}^\prime_{\rm q} \Delta^\prime_{sd} \Delta^\prime_{bd} \Delta^\prime_{bs} = I^6_{\rm d} \xi^2_u \xi^2_c \xi^2_t  {\cal J}^{}_{\rm q} \Delta^{}_{sd} \Delta^{}_{bd} \Delta^{}_{bs} \; .
\end{eqnarray}
In a similar way one may derive the Naumov-like relation for the RG evolution of charged leptons and Dirac neutrinos, namely,
\begin{eqnarray}
{\cal J}^\prime_\ell \Delta^\prime_{21} \Delta^\prime_{31} \Delta^\prime_{32} = I^6_\nu \zeta^2_e \zeta^2_\mu \zeta^2_\tau {\cal J}^{}_\ell \Delta^{}_{21} \Delta^{}_{31} \Delta^{}_{32} \; ,
\end{eqnarray}
with $\Delta^\prime_{ij} \equiv y^2_i - y^2_j$ and $\Delta^{}_{ij} = y^{2}_i - y^{2}_j$ (for $ij = 21, 31, 32$). Note that Eq. (19) has been derived in Ref. \cite{Xing:2017mkx}
by following a different approach in the MSSM and 2HDM frameworks.

Then, we proceed to explore other possible relations between the quark flavor mixing parameters at the electroweak scale $\Lambda^{}_{\rm EW}$ and those at a superhigh-energy scale $\Lambda$. Starting with the identity $H^\prime_{\rm d} = I^2_{\rm d} T^{}_{\rm d} H^{}_{\rm d} T^{}_{\rm d}$, we further adopt the top-dominance approximation, namely, $T^{}_{\rm d} \approx {\rm diag}\{1, 1, \xi^{}_t\}$. In this approximation, it has been demonstrated that the most convenient parametrization of the CKM matrix is~\cite{FX1, FX2}
\begin{eqnarray}
V = \left(\begin{matrix} c^{}_{\rm u} & s^{}_{\rm u} & 0 \cr -s^{}_{\rm u} & c^{}_{\rm u} & 0 \cr 0 & 0 & 1 \end{matrix}\right) \cdot \left(\begin{matrix} e^{{\rm i}\phi^{}_{\rm q}} & 0 & 0 \cr 0 & c & s \cr 0 & -s & c \end{matrix}\right)  \cdot \left(\begin{matrix} c^{}_{\rm d} & -s^{}_{\rm d} & 0 \cr s^{}_{\rm d} & c^{}_{\rm d} & 0 \cr 0 & 0 & 1 \end{matrix}\right)\; ,
\end{eqnarray}
where $c^{}_{\rm u} \equiv \cos \theta^{}_{\rm u}$ and $s^{}_{\rm u} \equiv \sin \theta^{}_{\rm u}$ have been defined and likewise for the other two mixing angles $\theta$ and $\theta^{}_{\rm d}$, and $\phi^{}_{\rm q}$ is the CP-violating phase. The parametrization in Eq.~(20) actually represents three sequential rotations in the flavor space $V = R^{}_{12}(\theta^{}_{\rm u}) \cdot U^{}_{\phi^{}_{\rm q}} \cdot R^{}_{23}(\theta) \cdot R^{\rm T}_{12}(\theta^{}_{\rm d})$, where $U^{}_{\phi^{}_{\rm q}} \equiv {\rm diag}\{e^{{\rm i}\phi^{}_{\rm q}}, 1, 1\}$ and $R^{}_{ij}(\theta)$ denotes the rotation in the $i$-$j$ plane with the rotation angle $\theta$ (for $ij = 12, 23$). Since $T^{}_{\rm d} \approx {\rm diag}\{1, 1, \xi^{}_t\}$ in the top-dominance approximation commutes with the rotation matrix $R^{}_{12}(\theta^{}_{\rm u}) U^{}_{\phi^{}_{\rm q}}$, we can obtain
\begin{eqnarray}
\widetilde{H}^\prime_{\rm d} &\equiv& \left[R^{}_{12}(\theta^{}_{\rm u}) U^{}_{\phi^{}_{\rm q}}\right]^\dagger \cdot H^\prime_{\rm d} \cdot \left[R^{}_{12}(\theta^{}_{\rm u}) U^{}_{\phi^{}_{\rm q}}\right] \nonumber \\
&=& I^2_{\rm d} T^{}_{\rm d} \left[ R^{}_{23}(\theta) R^{\rm T}_{12}(\theta^{}_{\rm d}) \cdot D^2_{\rm d} \cdot R^{}_{12}(\theta^{}_{\rm d}) R^{\rm T}_{23}(\theta) \right] T^{}_{\rm d} \; ,
\end{eqnarray}
which is a real and symmetric matrix and can be diagonalized by the following orthogonal transformation with three rotation angles $\{\widehat{\theta}^{}_{\rm u}, \widehat{\theta}, \widehat{\theta}^{}_{\rm d}\}$, i.e.,
\begin{eqnarray}
\widetilde{H}^\prime_{\rm d} = \left[R^{}_{12}(\widehat{\theta}^{}_{\rm u}) R^{}_{23}(\widehat{\theta}) R^{\rm T}_{12}(\widehat{\theta}^{}_{\rm d})\right] \cdot D^{\prime 2}_{\rm d} \cdot \left[R^{}_{12}(\widehat{\theta}^{}_{\rm u}) R^{}_{23}(\widehat{\theta}) R^{\rm T}_{12}(\widehat{\theta}^{}_{\rm d})\right]^{\rm T} \; .
\end{eqnarray}
Furthermore, $H^\prime_{\rm d}$ itself can be diagonalized via $H^\prime_{\rm d} = V^\prime D^{\prime 2}_{\rm d} V^{\prime \dagger}$, where the CKM matrix $V^\prime$ at the electroweak scale $\Lambda^{}_{\rm EW}$ can be parametrized in terms of $\{\theta^\prime_{\rm u}, \theta^\prime, \theta^\prime_{\rm d}, \phi^\prime_{\rm q}\}$ in the same way as in Eq.~(20). With the help of Eqs.~(21) and (22), we can recognize $V^\prime = \left[R^{}_{12}(\theta^{}_{\rm u}) U^{}_{\phi^{}_{\rm q}}\right] \cdot \left[R^{}_{12}(\widehat{\theta}^{}_{\rm u}) R^{}_{23}(\widehat{\theta}) R^{\rm T}_{12}(\widehat{\theta}^{}_{\rm d})\right]$ and then obtain
\begin{eqnarray}
R^{}_{12}(\theta^{}_{\rm u}) U^{}_{\phi^{}_{\rm q}} R^{}_{12}(\widehat{\theta}^{}_{\rm u}) = \left(\begin{matrix} e^{{\rm i}\varphi^{}_1} & 0 & 0 \cr 0 & e^{{\rm i}\varphi^{}_2 } & 0 \cr 0 & 0 & 1\end{matrix}\right)R^{}_{12} (\theta^\prime_{\rm u}) U^{}_{\phi^\prime_{\rm q}} \; ,
\end{eqnarray}
where $\varphi^{}_1$ and $\varphi^{}_2$ are the unphysical phases that can be absorbed by redefining the up-type quark fields. In addition, the other two mixing angles are given by $\theta^\prime = \widehat{\theta}$ and $\theta^\prime_{\rm d} = \widehat{\theta}^{}_{\rm d}$. In order to extract $\theta^\prime_{\rm u}$ and $\phi^\prime_{\rm q}$ from Eq.~(23), one has to write down the explicit forms of the matrices on both sides, and identify the corresponding matrix elements. To be more explicit, we have
\begin{eqnarray}
\sin \theta^\prime_{\rm u} &=& \left|\sin \theta^{}_{\rm u} \cos \widehat{\theta}^{}_{\rm u} + \cos \theta^{}_{\rm u} \sin \widehat{\theta}^{}_{\rm u} e^{{\rm i}\phi^{}_{\rm q}}\right| \; , \nonumber \\
\cos \theta^\prime_{\rm u} &=& \left|\cos \theta^{}_{\rm u} \cos \widehat{\theta}^{}_{\rm u} - \sin \theta^{}_{\rm u} \sin \widehat{\theta}^{}_{\rm u} e^{{\rm i}\phi^{}_{\rm q}}\right| \; ;
\end{eqnarray}
and
\begin{eqnarray}
\varphi^{}_1 &=& \arg\left[\sin \theta^{}_{\rm u} \cos \widehat{\theta}^{}_{\rm u} + \cos \theta^{}_{\rm u} \sin \widehat{\theta}^{}_{\rm u} e^{{\rm i}\phi^{}_{\rm q}}\right] \; , \nonumber \\
\varphi^{}_2 &=& \arg\left[\cos \theta^{}_{\rm u} \cos \widehat{\theta}^{}_{\rm u} - \sin \theta^{}_{\rm u} \sin \widehat{\theta}^{}_{\rm u} e^{{\rm i}\phi^{}_{\rm q}}\right] \; ,
\end{eqnarray}
and $\phi^\prime_{\rm q} = \phi^{}_{\rm q} - \varphi^{}_1 - \varphi^{}_2$. Therefore, we have established the relationship between the flavor mixing parameters $\{\theta^\prime_{\rm u}, \theta^\prime, \theta^\prime_{\rm d}, \phi^\prime_{\rm q}\}$ at $\Lambda^{}_{\rm EW}$ and $\{\theta^{}_{\rm u}, \theta, \theta^{}_{\rm d}, \phi^{}_{\rm q}\}$ at $\Lambda$ in an implicit way. Then it is straightforward to verify
\begin{eqnarray}
\sin \phi^\prime_{\rm q} \sin 2\theta^\prime_{\rm u} &=& 2{\rm Im}\left[e^{{\rm i}\phi^\prime_{\rm q}} \sin\theta^\prime_{\rm u} \cos\theta^\prime_{\rm u}\right] \nonumber \\
&=& 2{\rm Im} \left[\left(\sin \theta^{}_{\rm u} \cos \widehat{\theta}^{}_{\rm u} e^{{\rm i}\phi^{}_{\rm q}}  + \cos \theta^{}_{\rm u} \sin \widehat{\theta}^{}_{\rm u} \right) \left(\cos \theta^{}_{\rm u} \cos \widehat{\theta}^{}_{\rm u} - \sin \theta^{}_{\rm u} \sin \widehat{\theta}^{}_{\rm u} e^{-{\rm i}\phi^{}_{\rm q}}\right)\right] ~~~\quad \nonumber \\
&=& \sin\phi^{}_{\rm q}  \sin 2\theta^{}_{\rm u} \; ,
\end{eqnarray}
which is analogous to the Toshev relation in Eq.~(2) for neutrino oscillations in matter with the standard parametrization of the PMNS matrix~\cite{Toshev}.

Analogously, there is no doubt that the Toshev-like relation
\begin{eqnarray}
\sin \phi^\prime_\ell \sin 2\theta^\prime_l = \sin \phi^{}_\ell \sin 2\theta^{}_l
\end{eqnarray}
also holds in the RG evolution of charged-leptons and Dirac neutrinos from $\Lambda$
down to $\Lambda^{}_{\rm EW}$, if a similar parametrization as
that in Eq.~(20) is adopted for the PMNS matrix $U^\prime(\theta^\prime_l, \theta^\prime, \theta^\prime_\nu, \phi^\prime_\ell)$ at $\Lambda^{}_{\rm EW}$ or $U(\theta^{}_l, \theta, \theta^{}_\nu, \phi^{}_\ell)$ at $\Lambda$ and the tau-dominance approximation $T^{}_\nu \approx {\rm diag}\{1, 1, \zeta^{}_\tau\}$ is made. Although the RG equations of quark and lepton flavor mixing parameters have been extensively studied in the literature~\cite{Zhou}, we stress that such Toshev-like relations are derived here for the first time.

Some useful remarks are in order. First, the connection between the mixing parameters $\{\theta^{}_{\rm u}, \theta, \theta^{}_{\rm d}, \phi^{}_{\rm q}\}$ in Eq.~(20) and those $\{\vartheta^{}_{12}, \vartheta^{}_{13}, \vartheta^{}_{23}, \delta^{}_{\rm q}\}$ in the standard parametrization of the CKM matrix~\cite{PDG} can be found by identifying the moduli of four independent matrix elements and the Jarlskog invariant. More explicitly, we have~\cite{Fritzsch:1997st}
\begin{eqnarray}
\sin \vartheta^{}_{13} &=& \sin \theta^{}_{\rm u} \sin \theta \; , \nonumber \\
\sin \vartheta^{}_{23} &=& \frac{\cos \theta^{}_{\rm u} \sin \theta}{\sqrt{1 - \sin^2 \theta^{}_{\rm u} \sin^2 \theta}} \; , \nonumber \\
\tan \vartheta^{}_{12} &=& \frac{\left|\sin \theta^{}_{\rm u} \cos \theta^{}_{\rm d} \cos \theta - \cos \theta^{}_{\rm u} \sin \theta^{}_{\rm d} e^{{\rm i}\phi^{}_{\rm q}}\right|}{\left|\sin \theta^{}_{\rm u} \sin \theta^{}_{\rm d} \cos \theta + \cos \theta^{}_{\rm u} \cos \theta^{}_{\rm d} e^{{\rm i}\phi^{}_{\rm q}} \right|} \; , \nonumber \\
\sin \delta^{}_{\rm q} &=& \frac{\sin \theta^{}_{\rm u} \cos \theta^{}_{\rm u} \sin \theta^{}_{\rm d} \cos \theta^{}_{\rm d} \sin^2 \theta \cos \theta  \sin \phi^{}_{\rm q}}{\sin \vartheta^{}_{12} \cos \vartheta^{}_{12} \sin \vartheta^{}_{23} \cos \vartheta^{}_{23} \sin \vartheta^{}_{13} \cos^2 \vartheta^{}_{13}} \; .
\end{eqnarray}
Then, it should be noticed that the Toshev-like relation in Eq.~(26) for quarks has been derived in the flavor basis where the up-type quark Yukawa coupling matrix $Y^{}_{\rm u}$ is diagonal, so the mixing angle $\theta^{}_{\rm u}$ is just introduced to parametrize the CKM matrix and not necessarily coming from the up-type quark sector. However, in specific models of quark masses and flavor mixing~\cite{Fritzsch:1999ee}, $\theta^{}_{\rm u}$ and $\theta^{}_{\rm d}$ do arise from the unitary matrices diagonalizing the Yukawa coupling matrices of up- and down-type quarks, respectively. Therefore, after transforming into the diagonal basis of up-type quarks, the rotation angle $\theta^{}_{\rm u}$ appears in the CKM matrix. Similar arguments are also applicable to the lepton sector.

In general, the Naumov- and Toshev-like relations derived for the RG running of quark and lepton flavor mixing parameters should be useful for investigating the flavor models of quarks and leptons at a superhigh energy scale, given the experimental measurements at the low-energy scale. Although two such relations may not be sufficient to completely solve the RG equations, both of them are related to the CP-violating phase and thus lead to an interesting connection between low- and high-energy CP violation. Furthermore, they can be implemented to greatly simply the approximate analytical solutions to the RG equations.

\section{Summary}

The profundity and usefulness of RG equations can never be overemphasized, and this powerful tool has found many important applications in a number of different fields of modern physics~\cite{Zhou}. Recently, the RG-like equations have been derived in Refs.~\cite{Kuo} and \cite{XZZ} and implemented to the study of matter effects on neutrino oscillations, from which the previously known Naumov and Toshev relations have been rediscovered. We are therefore motivated to investigate if such kinds of relations exist for the RG evolution of fermion masses and flavor mixing parameters from a superhigh-energy scale down to the electroweak scale. It turns out that such Naumov- and Toshev-like relations {\it do} exist, as we have shown in this paper.

Though we work in a minimal extension of the SM with three massive Dirac neutrinos, in which the RG running effects are usually small, similar results are also obtainable in the MSSM and 2HDM cases only if the top- and tau-dominance approximations can be made. Since the Naumov- and Toshev-like relations directly connect the low- and high-energy flavor parameters, they definitely provide an interesting and transparent way to understand the RG-induced quantum corrections in this respect.

\vspace{0.6cm}

This work is supported in part by the National Natural Science Foundation of China under grant No. 11775231 (ZZX) and grant No. 11775232 (SZ), the National Youth Thousand Talents Program (SZ), and the CAS Center for Excellence in Particle Physics (SZ).



\begin{thebibliography}{99}

\bibitem{Zhou}
  T.~Ohlsson and S.~Zhou,
  ``Renormalization group running of neutrino parameters,''
  Nature Commun.\ {\bf 5}, 5153 (2014)
  [arXiv:1311.3846].

\bibitem{Kuo}
  S.~H.~Chiu and T.~K.~Kuo,
  ``Features of Neutrino Mixing,''
  Phys.\ Rev.\ D {\bf 97}, 055026 (2018)
  [arXiv:1712.08487].

\bibitem{XZZ}
  Z.~z.~Xing, S.~Zhou and Y.~L.~Zhou,
  ``Renormalization-Group Equations of Neutrino Masses and Flavor Mixing Parameters in Matter,''
  JHEP {\bf 1805}, 015 (2018)
  [arXiv:1802.00990].

\bibitem{Naumov}
  V.~A.~Naumov,
  ``Three neutrino oscillations in matter, CP violation and topological phases,''
  Int.\ J.\ Mod.\ Phys.\ D {\bf 1}, 379 (1992).

\bibitem{Toshev}
  S.~Toshev,
  ``On T violation in matter neutrino oscillations,''
  Mod.\ Phys.\ Lett.\ A {\bf 6}, 455 (1991).

\bibitem{PDG}
  C.~Patrignani {\it et al.} [Particle Data Group],
  ``Review of Particle Physics,''
  Chin.\ Phys.\ C {\bf 40}, 100001 (2016).

\bibitem{Harrison}
  P.~F.~Harrison and W.~G.~Scott,
  ``CP and T violation in neutrino oscillations and invariance of Jarlskog's determinant to matter effects,''
  Phys.\ Lett.\ B {\bf 476}, 349 (2000)
  [hep-ph/9912435].

\bibitem{Xing2001}
  Z.~z.~Xing,
  ``New formulation of matter effects on neutrino mixing and CP violation,''
  Phys.\ Lett.\ B {\bf 487}, 327 (2000)
  [hep-ph/0002246].

\bibitem{Xing2001x}
  Z.~z.~Xing,
  ``Sum rules of neutrino masses and CP violation in the four neutrino mixing scheme,''
  Phys.\ Rev.\ D {\bf 64}, 033005 (2001)
  [hep-ph/0102021].

\bibitem{FX1}
  H.~Fritzsch and Z.~z.~Xing,
  ``Flavor symmetries and the description of flavor mixing,''
  Phys.\ Lett.\ B {\bf 413}, 396 (1997)
  [hep-ph/9707215].

\bibitem{FX2}
  H.~Fritzsch and Z.~z.~Xing,
  ``Large leptonic flavor mixing and the mass spectrum of leptons,''
  Phys.\ Lett.\ B {\bf 440}, 313 (1998)
  [hep-ph/9808272].

\bibitem{XZ}
  Z.~z.~Xing and S.~Zhou,
  ``Neutrinos in particle physics, astronomy and cosmology,''
  Springer-Verlag, Berlin Heidelberg (2011).

\bibitem{Lindner:2005as}
  M.~Lindner, M.~Ratz and M.~A.~Schmidt,
  ``Renormalization group evolution of Dirac neutrino masses,''
  JHEP {\bf 0509}, 081 (2005)
  [hep-ph/0506280].

\bibitem{Antusch:2003kp}
  S.~Antusch, J.~Kersten, M.~Lindner and M.~Ratz,
  ``Running neutrino masses, mixings and CP phases: Analytical results and phenomenological consequences,''
  Nucl.\ Phys.\ B {\bf 674}, 401 (2003)
  [hep-ph/0305273].

\bibitem{Antusch:2005gp}
  S.~Antusch, J.~Kersten, M.~Lindner, M.~Ratz and M.~A.~Schmidt,
  ``Running neutrino mass parameters in see-saw scenarios,''
  JHEP {\bf 0503}, 024 (2005)
  [hep-ph/0501272].

\bibitem{Mei:2005qp}
  J.~w.~Mei,
  ``Running neutrino masses, leptonic mixing angles and CP-violating phases: From M(Z) to Lambda(GUT),''
  Phys.\ Rev.\ D {\bf 71}, 073012 (2005)
  [hep-ph/0502015].

\bibitem{Zhao}
  Z.~z.~Xing and Z.~h.~Zhao,
  ``A review of $\mu$-$\tau$ flavor symmetry in neutrino physics,''
  Rept.\ Prog.\ Phys.\  {\bf 79}, 076201 (2016)
  [arXiv:1512.04207].

\bibitem{Minkowski}
  P.~Minkowski,
  ``$\mu \to e\gamma$ at a Rate of One Out of $10^{9}$ Muon Decays?,''
  Phys.\ Lett.\  {\bf 67B}, 421 (1977).

\bibitem{Yanagida}
  T.~Yanagida,
  ``Horizontal Symmetry And Masses Of Neutrinos,''
  Conf.\ Proc.\ C {\bf 7902131}, 95 (1979).

\bibitem{Glashow}
  S.~L.~Glashow,
  ``The Future of Elementary Particle Physics,''
  NATO Sci.\ Ser.\ B {\bf 61}, 687 (1980).

\bibitem{Gell-Mann}
  M.~Gell-Mann, P.~Ramond and R.~Slansky,
  ``Complex Spinors and Unified Theories,''  Conf.\ Proc.\ C {\bf 790927}, 315 (1979)  [arXiv:1306.4669].

\bibitem{Mohapatra}
  R.~N.~Mohapatra and G.~Senjanovic,
  ``Neutrino Mass and Spontaneous Parity Violation,''
  Phys.\ Rev.\ Lett.\  {\bf 44}, 912 (1980).

\bibitem{Xing:2017mkx}
  Z.~z.~Xing, D.~Zhang and J.~y.~Zhu,
  ``The $\mu-\tau$ reflection symmetry of Dirac neutrinos and its breaking effect via quantum corrections,''
  JHEP {\bf 1711}, 135 (2017)
  [arXiv:1708.09144].

\bibitem{Xing:2011aa}
  Z.~z.~Xing, H.~Zhang and S.~Zhou,
  ``Impacts of the Higgs mass on vacuum stability, running fermion masses and two-body Higgs decays,''
  Phys.\ Rev.\ D {\bf 86}, 013013 (2012)
  [arXiv:1112.3112].

\bibitem{Jarlskog:1985ht}
  C.~Jarlskog,
  ``Commutator of the Quark Mass Matrices in the Standard Electroweak Model and a Measure of Maximal CP Violation,''
  Phys.\ Rev.\ Lett.\  {\bf 55}, 1039 (1985).

\bibitem{Fritzsch:1997st}
  H.~Fritzsch and Z.~z.~Xing,
  ``On the parametrization of flavor mixing in the standard model,''
  Phys.\ Rev.\ D {\bf 57}, 594 (1998)
  [hep-ph/9708366].

\bibitem{Fritzsch:1999ee}
  H.~Fritzsch and Z.~z.~Xing,
  ``Mass and flavor mixing schemes of quarks and leptons,''
  Prog.\ Part.\ Nucl.\ Phys.\  {\bf 45}, 1 (2000)
  [hep-ph/9912358].
\end{thebibliography}
\end{document}